\documentclass[12pt]{article}
\usepackage{graphicx}
\usepackage{dcolumn}
\usepackage{bm}
\usepackage{amsmath}
\usepackage{amssymb}
\usepackage{color}
\usepackage{hyperref} 
\title{
Matters Arising: \\
Overcoming a challenge for Bohmian mechanics}
\author{
Hrvoje Nikoli\'c \\
Theoretical Physics Division, Rudjer Bo\v{s}kovi\'{c} Institute, \\
P.O.B. 180, HR-10002 Zagreb, Croatia \\
{\normalsize e-mail: hnikolic@irb.hr} 
}
\date{\today}
\begin{document}
\maketitle


Bohmian mechanics \cite{book-hol} is an alternative formulation of quantum mechanics 
in terms of particles that have well defined deterministic trajectories. 
Recently, an interesting challenge \cite{nature25} has been posed for Bohmian mechanics by studying a 
quantum system
in which the motion of particles cannot be associated only with the gradient of phase of the wave function.
We point out that, in general, Bohmian velocity is defined by the continuity equation, which does not always 
lead to velocity depending only on the phase gradient. By constructing the appropriate velocity explicitly,
we overcome the challenge posed in \cite{nature25}.

More specifically, in \cite{nature25}, the authors studied the population transfer between two coupled waveguides
and found that there is associated motion of particles that cannot be explained by the Bohmian velocity
\begin{equation}\label{B1}
 {\bf v}({\bf x},t)=\frac{\hbar}{m} \mbox{\boldmath $\nabla$} \phi({\bf x},t), 
\end{equation}
where $\phi({\bf x},t)$ is the phase of the wave function $\psi({\bf x},t)=\sqrt{\rho({\bf x},t)}\, e^{i\phi({\bf x},t)}$,
and $m$ is the mass of the particle.
Since they posed this as an interesting challenge for Bohmian mechanics, in this paper we overcome their challenge
by explaining why the usual Bohmian velocity (\ref{B1}) does not work in their case, and find
a modified Bohmian velocity that works.    
   
Our key point is that, in Bohmian mechanics, the velocity is not always given by the formula (\ref{B1}).
In general, the velocity in Bohmian mechanics is found from the requirement that it satisfies the continuity equation,
which in the one-particle case takes the form
\begin{equation}\label{cont1}
 \frac{\partial\rho}{\partial t} + \mbox{\boldmath $\nabla$} (\rho {\bf v}) =0 .
\end{equation}
This leads to the velocity (\ref{B1}) when $\psi$ satisfies a Schr\"odinger equation of the form 
\begin{equation}\label{sch1}
 \frac{-\hbar^2\mbox{\boldmath $\nabla$}^2}{2m}\psi +V\psi = -i\hbar\partial_t\psi .
\end{equation}
However, for a Schr\"odinger-like equation which is not of that form, the formula for the velocity may be different.
A very general method for finding the appropriate Bohmian velocity in general quantum systems has been developed in 
\cite{nikferm}. 

In \cite{nature25}, the authors model the coupled waveguides by two coupled time-independent Schr\"odinger equations
\begin{eqnarray}\label{sch2}
& E\psi_m = -\displaystyle\frac{\hbar^2}{2m}\frac{d^2\psi_m}{dx^2} + V_0\psi_m + \hbar J_0 (\psi_a - \psi_m), &
\nonumber \\
& E\psi_a = -\displaystyle\frac{\hbar^2}{2m}\frac{d^2\psi_a}{dx^2} + V_0\psi_a + \hbar J_0 (\psi_m - \psi_a), &
\end{eqnarray}
where $\psi_m$ and $\psi_a$ represent the wave functions in the main waveguide and the auxiliary waveguide, respectively,
while $V_0$ and $J_0$ are constants. (They actually made a typo by missing the square in $\hbar^2$, but this did not affect
their results.)
They find that appropriate solutions of these coupled equations take the form 
\begin{equation}\label{wf}
 \psi_m(x) \propto \cos(k_1x)e^{ik_2x}, \;\;\; \psi_a(x) \propto \sin(k_1x)e^{ik_2x} .
\end{equation}
Note that each of the equations in (\ref{sch2}) involves {\em two} wave functions, $\psi_m$ and $\psi_a$,
so neither of the equations is of the form (\ref{sch1}) which involves only one wave function $\psi$. 
Thus, {\it a priori}, there is no reason to expect that the appropriate Bohmian velocity 
associated with (\ref{sch2}) should be given by (\ref{B1}).

Our goal now is to find the Bohmian velocity consistent with wave functions (\ref{wf}). Since the wave functions depends 
only on $x$, the continuity equation (\ref{cont1}) reduces to
\begin{equation}\label{cont2}
\partial_x(\rho v_x)= 0 ,  
\end{equation}
where $\rho=|\psi|^2$, for $\psi=\psi_m$ in the main waveguide and $\psi=\psi_a$ in the auxiliary waveguide.
The continuity equation (\ref{cont2}) is satisfied if and only if the velocity is of the form
\begin{equation}\label{B2}
 v_x(x)=\frac{c}{\rho(x)} ,
\end{equation}
where $c$ is a constant. Restricting our attention to real $k_1$ and $k_2$, from (\ref{wf}) we have  
\begin{equation}\label{dens}
\rho_m(x)\propto \cos^2(k_1x), \;\;\; \rho_a(x)\propto \sin^2(k_1x) ,
\end{equation}
so we see that the velocity (\ref{B2}) has a nontrivial dependence on $x$, not given by the formula (\ref{B1}).
The constant $c$ can be fixed by requiring \cite{nikferm} that the {\em average} velocity is the same as the average of the 
velocity (\ref{B1}) 
\begin{equation}
 \int_0^L dx\, \rho(x) v_x(x) = \int_0^L dx\, \rho(x) \frac{\hbar}{m}\partial_x \phi(x) ,
\end{equation}
where $L$ is the length of the waveguide. This requirement leads to the final result
\begin{equation}\label{B3}
 v_x(x)=\frac{1}{L\rho(x)} \frac{\hbar k_2}{m} ,
\end{equation}
for the wave function normalized so that $\int_0^L dx\, \rho(x)=1$. 
This velocity can be thought of as a modification and generalization of (\ref{B1}). In particular, 
if $\rho$ is a constant $\rho=1/L$, then (\ref{B3}) reduces to $v_x=\hbar k_2/m$, which coincides 
with the velocity given by (\ref{B1}).   
    
Finally note that the Bohmian velocity $v_x$ differs from the speed $v$ measured in \cite{nature25}. Indeed, 
the speed $v$ in \cite{nature25} does not have a direction and cannot be interpreted as a velocity, 
while the Bohmian $v_x$ is a velocity
directed in the $x$-direction. It does not pose any contradiction between our results and those of \cite{nature25},
because it is a common feature of Bohmian velocities that they cannot be measured directly. Nevertheless, the 
Bohmian velocity has an explanatory power, in the sense that it can explain the densities (\ref{dens})
in terms of motions associated with ``hidden'' variables.    

Acknowledgment: The author is grateful to N. Bili\'c for drawing attention to the work \cite{nature25}.

\end{document}